# Design and analysis of deployable clustered tensegrity cable domes

Shuo MA[a], Muhao CHEN[b,]*, Xingfei YUAN[c], Robert E. SKELTON[b]

*Department of Aerospace Engineering, Texas A&M University, College Station
701 H.R. Bright Building 3141, College Station, TX 77843, USA
muhaochen@tamu.edu (M. CHEN)

[a] College of Civil Engineering, Zhejiang University of Technology, Hangzhou, 310014, China
[b] Department of Aerospace Engineering, Texas A&M University, College Station, TX, 77840, USA
[c] College of Civil Engineering and Architecture, Zhejiang University, Hangzhou, 310058, China

## Abstract

This study presents the design and analysis of deployable cable domes based on the clustered tensegrity structures (CTS). In this paper, the statics and dynamics equations of the CTS are first given. Using a traditional Levy cable dome as an example, we show the approach to modify the Levy dome to a deployable CTS one. The strings to be clustered are determined by the requirement of prestress mode and global stability. The deployment trajectory is proposed by changing the deployment ratio (the ratio between the radius of the inner and outer rings of the cable dome). Then, the quasi-static and dynamic deployment of clustered tensegrity dome is studied. Results show that the proposed CTS cable dome always has one prestress mode and is globally stable in its deployment trajectory. In the deployment process analysis, the dynamics show that the system's dynamic response differs from the quasi-static simulation as the actuation speed increases. That is, for a fast deployment process, quasi-static simulation is not accurate enough. The dynamics effects of the deployment must be considered. The developed approaches can also be used for the design and analysis of various kinds of CTS.

## 1. Introduction

A cable dome usually consists of an inner tension ring and an outer compression ring. The two rings are stabilized by ridge strings, diagonal strings, hoop strings, and vertical struts [1]. The certain stiffness of the cable dome is achieved by tuning and adjusting the prestress in the cables. The first cable dome was built for the roof of the Olympic gymnastics hall (diameter 120m circular plan) by Geiger and was employed by the Fencing hall (diameter 90m circular plan) in Seoul in 1986 [2]. Since then, due to the many nice properties of cable domes, such as lightweight, architectural aesthetics, and large special span, cable domes have become a popular design option for roof-like structures, including arenas, stadiums, and sports centers. To list a few, Redbird Arena (long axis 91m and short axis 77m elliptical plan), the first Geiger cable dome in North America, is a 10,500 multi-purpose arena built in 1989 located in Normal, Illinois, US [3]. Georgia Dome (long axis 240m and short axis 192m elliptical plan) can host 80,000 people was constructed in Atlanta, Georgia, US in 1992 [4]. Tropicana Field, also known as the Suncoast Dome (diameter 210m circular plan), has a capacity of 42,735 seats, was built in 1990, St. Petersburg, Florida, US [5]. La Plata Stadium (field size 105m × 70 m), which can hold 53,000 people, was opened in 2003, La Plata, Argentina. Tao-Yuan County Arena (diameter 82m circular plan), which has 15,000 seats, was completed in 1993 in Taoyuan, Taiwan, China [6]. New

Area Technology Communication Center (field size 143m × 140m, cable dome roof size diameter 24m circular plan), Wuxi, Jiangsu, China. Taiyuan Coal Exchange Center (field size 260m in diameter, cable dome roof size diameter 36m circular plan), Taiyuan, Shanxi, China.

After decades of study, a few types of research have been conducted on cable domes designs in many aspects. For example, Heitel and Fu studied the form-finding and structural optimizations for tensile using Grasshopper in Rhino [7]. Guo and Jiang presented an approach to find the feasible prestress for cable-struts structures with and without loads. Chen et al. showed a robustness analysis for spatial flexible cable-strut tensile structures in span, structure complexity, redundancy, construction deviations based on $H_\infty$ theory [8]. Krishnan demonstrated the influence of geometric parameters and cable prestressing forces on prestressed radial-type cable domes [9]. Kmet and Mojdis described a Levy form-based adaptive cable dome, altering stiffness configuration and stress properties to adapt its behaviours to its current loading conditions [10]. However, almost all the built cable domes are rigid ones. In fact, most of the cable domes have the ability to be deployable by tuning the length of certain strings. One of the best properties of deployability is that one can adjust the sunlight coming through the inner space as a concept of green/low-carbon architecture. This paper focuses on the analysis of a deployable cable dome.

By the configuration of the cable domes (made of bars and strings), we know it is consistent with the definition of tensegrity. Tensegrity is a coined word of tension and integrity by Buckminster Fuller for the art form by Ioganson (1921) and Snelson (1948) [11]. The tensegrity system is a stable network of compressive members (bars/struts) and tensile members (cables/strings). After decades of study, tensegrity has shown its many advantages: 1). Lightweight solutions to five fundamental problems in engineering mechanics (compression, tension, torsion, cantilever, and simply supported) [12-14]. 2). The deformation of the structure members is one-dimensional, which provides more accurate models [15, 16]. 3). One can achieve large morphing objectives by tuning the length of strings [17, 18]. 4). The strings in the structure can be used to absorb energy [19, 20]. 5). Promoting the integration of structure and control design since the structure members can be both actuators and sensors [21-23].

In the classic formulation of the tensegrity governing equations, strings are usually mathematically treated as separated ones divided at the structure nodes [24-26]. However, this is not a general approach. Because in the real constructions, two strings connected/clustered at the common node can be replaced by one string and a pulley at the common node to: 1.) reduce the number of strings, 2). reduce the number of sensors, actuators, and related devices, 3). simplify the process of assembly and prestress of strings. For tensegrity structures with clustered strings, it is called clustered tensegrity structures (CTS). Few studies have been conducted on this topic. For example, Kan et al. formulated an approach to analyze clustered tensegrity structures by generalized coordinates of the system [27]. Zhang et al. derived the finite element formulation for clustered tensegrity based on the co-rotational approach [28]. Kan et al. presented sliding cable elements for multibody system analysis by using the configuration of the attached rigid bodies as the generalized coordinates and investigated the deployment of clustered tensegrity [29]. Ali et al. presented a modified dynamic relaxation (DR) algorithm for static analysis and form-finding [30]. Moored and Bart-Smith demonstrated a clustered actuation of tensegrity structures and discussed prestress analysis, mechanism analysis, and stability of clustered structures [31]. This paper presented the design of a deployable tensegrity dome, allowing any of the two connected strings to be clustered, and studied the prestress mode and global stability and the deployment of clustered tensegrity both quasi-statically and dynamically.

This paper is organized as follows. Section 2 presents the nonlinear dynamics and statics equations (in terms of node coordinates, force densities, forces) of the CTS. Section 3 demonstrated the topology design, arrangement, and selection of the clustered strings based on the prestress mode and global stability of the tensegrity cable dome. Section 4 shows the analysis of the quasi-static and dynamics of the deployment of the cable dome. Section 5 summarizes the conclusions.

## 2. Statics and dynamics of the CTS

### 2.1 Assumptions and notations of the CTS

Under the following assumptions, the statics and dynamics of CTS are formulated: 1). The structural members are axially loaded, all structural members are connected by frictionless pin-joints. 2). Two strings connected at a common node can be clustered into one string, the force density in the clustered two strings is the same. 3). A string can never push along its length. Assume the CTS has $n_n$ number of nodes, the $X$-, $Y$-, and $Z$-coordinates of the $i$th node $i$ $(i = 1, 2, \cdots, n_n)$ can be labeled as $x_i$, $y_i$, and $z_i$: $\boldsymbol{n}_i = [x_i\ y_i\ z_i]^T$. Then, the nodal coordinate vector $\boldsymbol{n} \in \mathrm{R}^{3n_n}$ for the whole structure is:

$$\boldsymbol{n} = \left[\boldsymbol{n}_1^T\ \boldsymbol{n}_2^T\ \cdots \boldsymbol{n}_{n_n}^T\right]^T. \tag{1}$$

Let index matrices $\boldsymbol{E}_a \in \mathbb{R}^{3n_n \times n_a}$ and $\boldsymbol{E}_b \in \mathbb{R}^{3n_n \times n_b}$ be permutation matrix replacing $\boldsymbol{n}$ to free and constrained nodal coordinate vectors $\boldsymbol{n}_a$ and $\boldsymbol{n}_b$:

$$\boldsymbol{n}_a = \boldsymbol{E}_a^T \boldsymbol{n}, \boldsymbol{n}_b = \boldsymbol{E}_b^T \boldsymbol{n}. \tag{2}$$

Since $[\boldsymbol{E}_a\ \boldsymbol{E}_b]$ is an orthogonal matrix, we also have the following equation:

$$\boldsymbol{n} = \begin{bmatrix} \boldsymbol{E}_a^T \\ \boldsymbol{E}_b^T \end{bmatrix}^{-1} \begin{bmatrix} \boldsymbol{n}_a \\ \boldsymbol{n}_b \end{bmatrix} = [\boldsymbol{E}_a \quad \boldsymbol{E}_b] \begin{bmatrix} \boldsymbol{n}_a \\ \boldsymbol{n}_b \end{bmatrix}. \tag{3}$$

The bars and strings connection patterns are denoted by $\boldsymbol{C}_b \in R^{\beta \times n_n}$ and $\boldsymbol{C}_s \in R^{\alpha \times n_n}$, which are called the bar and string connectivity matrices. In this formulation, we use an integrated connectivity matrix $\boldsymbol{C}^T = \left[\boldsymbol{C}_b^T\ \boldsymbol{C}_s^T\right]$, the first $\beta$ rows are bars, and the last $\alpha$ rows are strings. The $i$th row of C, labeled as $\boldsymbol{C}_i = [\boldsymbol{C}]_{i,:}$ denotes the $i$th structure element, starting from node $j$ $(j = 1, 2, \cdots, n_n)$ to node $k$ $(k = 1, 2, \cdots, n_n)$, which satisfies:

$$[\boldsymbol{C}]_{im} = \begin{cases} -1 & m = j \\ 1 & m = k \\ 0 & m = \text{else} \end{cases}. \tag{4}$$

For easy deployability (i.e., use fewer motors to pull strings), some adjacent strings in the classical tensegrity structures can be replaced by a single string running over the string-string joints by pulleys. We call the single string a clustered string. To simplify the model, we assume the pulleys are negligibly small and frictionless. Suppose the number of elements in the traditional tensegrity structure is $n_e$, after clustering some strings, the number of clustered tensegrity structure becomes $n_{ec}$. We define a cluster matrix $\boldsymbol{S} \in \mathbb{R}^{n_{ec} \times n_e}$ to label which strings are clustered:

$$[\boldsymbol{S}]_{ij} = \begin{cases} 1, \text{ if the } i\text{th clustered element is composed of the } j\text{th classic element} \\ 0, \text{ otherwise} \end{cases}. \tag{5}$$

The structure element vector can be labeled by the nodal coordinates of its two sides, for example, the $m$th element vector $\boldsymbol{h}_m$ is given by:

$$\boldsymbol{h}_m = \boldsymbol{n}_k - \boldsymbol{n}_j = \boldsymbol{C}_m \otimes \mathbf{I}_3 \boldsymbol{n}, \tag{6}$$

and its length of can be calculated as:

$$\boldsymbol{l}_m = \|\boldsymbol{h}_m\| = [\boldsymbol{n}^T (\boldsymbol{C}_m^T \boldsymbol{C}_m) \otimes \mathbf{I}_3 \boldsymbol{n}]^{\frac{1}{2}}. \tag{7}$$

The structure member matrix $\boldsymbol{H} \in \mathbb{R}^{3 \times n_e}$, structure element length vector $\boldsymbol{l} \in \mathbb{R}^{n_e}$, and rest length vector $\boldsymbol{l}_0 \in \mathbb{R}^{n_e}$ can be written as:

$$\boldsymbol{H} = [\boldsymbol{h}_1 \quad \boldsymbol{h}_2 \quad \cdots \quad \boldsymbol{h}_{n_e}] = \boldsymbol{N}\boldsymbol{C}^{\boldsymbol{T}}, \tag{8}$$

$$\boldsymbol{l} = [l_1 \quad l_2 \quad \cdots \quad l_{n_e}]^T, \tag{9}$$

$$\boldsymbol{l}_0 = [l_{01} \quad l_{02} \quad \cdots \quad l_{0n_e}]^T, \tag{10}$$

where rest length is defined as the length of a structure element (bar or string) with no compression or tension. The CTS element length vector $\boldsymbol{l}_{\mathrm{c}} \in \mathbb{R}^{n_{ec}}$ is:

$$\boldsymbol{l}_c = \boldsymbol{S}\boldsymbol{l}. \tag{11}$$

Similarly, the CTS elements' rest length vector $\boldsymbol{l}_{0\mathrm{c}} \in \mathbb{R}^{n_{ec}}$ is:

$$\boldsymbol{l}_{0\mathrm{c}} = \boldsymbol{S}\boldsymbol{l}_0. \tag{12}$$

The cross-sectional area, secant modulus, tangent modulus of the $i$th element in TTS (traditional tensegrity structure, whose strings are independent ones, not clustered) are $A_i$, $E_i$, and $E_{ti}$. Suppose material density is $\rho$, the element mass $m_i$ satisfies $m_i = \rho A_i l_{0i}$. Denote the cross-sectional area vector, secant modulus vector, tangent modulus vector, and mass vector of the CTS as $\boldsymbol{A}_c$, $\boldsymbol{E}_c$, $\boldsymbol{E}_{tc}$, and $\boldsymbol{m}_{\mathrm{c}} \in \mathbb{R}^{n_{ec}}$:

$$\boldsymbol{A}_c = [A_{c1} \quad A_{c2} \quad \cdots \quad A_{cn_{ec}}]^T, \tag{13}$$

$$\boldsymbol{E}_c = [E_{c1} \quad E_{c2} \quad \cdots \quad E_{cn_{ec}}]^T, \tag{14}$$

$$\boldsymbol{E}_{tc} = \left[E_{ct1} \quad E_{ct2} \quad \cdots \quad E_{ctn_{ec}}\right]^T, \tag{15}$$

$$\boldsymbol{m}_c = [m_{c1} \quad m_{c2} \quad \cdots \quad m_{cn_{ec}}]^T = \rho \widehat{\boldsymbol{A}}_c \hat{\boldsymbol{l}}_{0c}, \tag{16}$$

where $\hat{\boldsymbol{v}}$ transforms vector $\boldsymbol{v}$ into a diagonal matrix, whose diagonal entries are vector $\boldsymbol{v}$ and elsewhere are zeros. Assume the property is consistent along a clustered string, the cross-sectional area, secant modulus, tangent modulus, and mass vector of the corresponding TTS is $\boldsymbol{A}$, $\boldsymbol{E}$, $\boldsymbol{E}_t$, and $\boldsymbol{m} \in \mathbb{R}^{n_e}$:

$$\boldsymbol{A} = [A_1 \quad A_2 \quad \cdots \quad A_{n_e}]^T = \boldsymbol{S}^T \boldsymbol{A}_c, \tag{17}$$

$$\boldsymbol{E} = [E_1 \quad E_2 \quad \cdots \quad E_{n_e}]^T = \boldsymbol{S}^T \boldsymbol{E}_c, \tag{18}$$

$$\boldsymbol{E}_t = [E_{t1} \quad E_{t2} \quad \cdots \quad E_{tn_e}]^T = \boldsymbol{S}^T \boldsymbol{E}_{tc}, \tag{19}$$

$$\boldsymbol{m} = [m_1 \quad m_2 \quad \cdots \quad m_{n_e}]^T = \rho \hat{\boldsymbol{A}} \boldsymbol{l}_0. \tag{20}$$

The internal force of the $i$th element in CTS is $t_{ci} = A_{ci}\sigma_{ci} = E_{ci}A_{ci}(l_{ci} - l_{0ci})/l_{0ci}$, so the internal force vector of CTS and TTS can be written as:

$$\boldsymbol{t}_c = \widehat{\boldsymbol{E}}_{\mathrm{c}} \widehat{\boldsymbol{A}}_{\mathrm{c}} \hat{\boldsymbol{l}}_{0\mathrm{c}}^{-1} (\boldsymbol{l}_c - \boldsymbol{l}_{0c}), \tag{21}$$

$$\boldsymbol{t} = \hat{\boldsymbol{E}} \hat{\boldsymbol{A}} \hat{\boldsymbol{l}}_0^{-1} (\boldsymbol{l} - \boldsymbol{l}_0) = \boldsymbol{S}^T \boldsymbol{t}_c. \tag{22}$$

Force density of the $i$th element in CTS is given by $x_{ci} = t_{ci}/l_{ci}$ the force density vector of CTS and TTS can be written as:

$$\boldsymbol{x}_c = \hat{\boldsymbol{l}}_c^{-1} \boldsymbol{t}_c = \widehat{\boldsymbol{E}}_c \widehat{\boldsymbol{A}}_c (\boldsymbol{l}_{0c}^{-1} - \boldsymbol{l}_c^{-1}), \tag{23}$$

$$\boldsymbol{x} = \hat{\boldsymbol{E}} \hat{\boldsymbol{A}} (\boldsymbol{l}_0^{-1} - \boldsymbol{l}^{-1}) = \boldsymbol{S}^T \boldsymbol{x}_c, \tag{24}$$

where $\boldsymbol{v}^{-1}$ represents a vector, whose entry is the reciprocal of its corresponding entry in $\boldsymbol{v}$. We should point out that Eqs. (21) and (22) can be used to compute force and force density vectors for either elastic or plastic materials using accurate secant modulus $\boldsymbol{E}$ of the materials.

## 2.2. Dynamics equation of CTS with constraints

The tensegrity dynamics with constraints can be given in the following vector form [32]:

$$\boldsymbol{E}_a^T(\boldsymbol{M}\ddot{\boldsymbol{n}} + \boldsymbol{D}\dot{\boldsymbol{n}} + \boldsymbol{K}\boldsymbol{n}) = \boldsymbol{w}_a, \tag{25}$$

where the external force vector is $\boldsymbol{w}_a = \boldsymbol{E}_a^T(\boldsymbol{f}_{ex} - \boldsymbol{g})$, $\boldsymbol{f}_{ex}$ is external forces on the structure nodes, and $\boldsymbol{g}$ is the gravity vector. The dynamic equation Eq. (25) has $\boldsymbol{n}_a$ rows corresponding to the free nodal coordinates. Rearrange the terms related to $\boldsymbol{n}_a$ in left side, we obtain:

$$\boldsymbol{M}_{aa}\ddot{\boldsymbol{n}}_a + \boldsymbol{D}_{aa}\dot{\boldsymbol{n}}_a + \boldsymbol{K}_{aa}\boldsymbol{n}_a = \boldsymbol{E}_a^T\boldsymbol{f}_{ex} - \boldsymbol{M}_{ab}\ddot{\boldsymbol{n}}_b - \boldsymbol{D}_{ab}\dot{\boldsymbol{n}}_b - \boldsymbol{K}_{ab}\boldsymbol{n}_b - \boldsymbol{E}_a^T\boldsymbol{g}, \tag{26}$$

where $\boldsymbol{M}_{aa}, \boldsymbol{D}_{aa}, \boldsymbol{K}_{aa}$ and $\boldsymbol{M}_{ab}, \boldsymbol{D}_{ab}, \boldsymbol{K}_{ab}$ are the mass matrices, damping matrices, stiffness matrices corresponding to free and constrained nodal coordinate, which satisfy:

$$\begin{aligned} \boldsymbol{M}_{aa} &= \boldsymbol{E}_a^T\boldsymbol{M}\boldsymbol{E}_a, \boldsymbol{M}_{ab} = \boldsymbol{E}_a^T\boldsymbol{M}\boldsymbol{E}_b \\ \boldsymbol{D}_{aa} &= \boldsymbol{E}_a^T\boldsymbol{D}\boldsymbol{E}_a, \boldsymbol{D}_{ab} = \boldsymbol{E}_a^T\boldsymbol{D}\boldsymbol{E}_b \ , \\ \boldsymbol{K}_{aa} &= \boldsymbol{E}_a^T\boldsymbol{K}\boldsymbol{E}_a, \boldsymbol{K}_{ab} = \boldsymbol{E}_a^T\boldsymbol{K}\boldsymbol{E}_b \end{aligned} \tag{27}$$

in which $\boldsymbol{M}, \boldsymbol{D}, \boldsymbol{K},$ and $\boldsymbol{g}$ are mass matrix, stiffness matrix, and gravity force:

$$\boldsymbol{M} = \frac{1}{6}\left(|\boldsymbol{C}|^{\mathrm{T}}\widehat{\boldsymbol{m}}|\boldsymbol{C}| + \left\lfloor|\boldsymbol{C}|^{\mathrm{T}}\widehat{\boldsymbol{m}}|\boldsymbol{C}|\right\rfloor\right)\otimes\boldsymbol{I}_3, \tag{28}$$

$$\boldsymbol{K} = \left(\boldsymbol{C}^{\mathrm{T}}\widehat{\boldsymbol{S}^T\boldsymbol{x}_c}\mathbf{C}\right)\otimes\boldsymbol{I}_3, \tag{29}$$

$$\boldsymbol{g} = \frac{g}{2}\left(|\boldsymbol{C}|^{\mathrm{T}}\boldsymbol{m}\right)\otimes[0\ 0\ 1]^{\mathrm{T}}, \tag{30}$$

in which mass vector $\boldsymbol{m} = \rho\widehat{\boldsymbol{A}}\left(\widehat{\boldsymbol{S}^T\boldsymbol{t}_c} + \widehat{\boldsymbol{E}}\widehat{\boldsymbol{A}}\right)^{-1}\widehat{\boldsymbol{E}}\widehat{\boldsymbol{A}}\boldsymbol{l}$ should be recalculated in every time step and used to renew the mass matrix. Note that the mass vector $\boldsymbol{m}$ in Eq. (28) and force density vector $\boldsymbol{x}_c$ in Eq. (29) should be recalculated by Eq. (20) and Eq. (23) in every time step when solving the dynamics equation. Take the total derivative of the nonlinear dynamics and keep the linear terms. We have the linearized dynamics equation of the CTS:

$$\boldsymbol{M}_{\boldsymbol{aa}}\mathrm{d}\ddot{\boldsymbol{n}}_{\boldsymbol{a}} + \boldsymbol{D}_{\boldsymbol{aa}}\mathrm{d}\dot{\boldsymbol{n}}_{\boldsymbol{a}} + \boldsymbol{K}_{\boldsymbol{Taa}}\mathrm{d}\boldsymbol{n}_{\boldsymbol{a}} = \boldsymbol{E}_a^T\mathrm{d}\boldsymbol{f}_{ex} - \boldsymbol{M}_{\boldsymbol{ab}}\mathrm{d}\ddot{\boldsymbol{n}}_{\boldsymbol{b}} - \boldsymbol{D}_{\boldsymbol{ab}}\mathrm{d}\dot{\boldsymbol{n}}_{\boldsymbol{b}} - \boldsymbol{K}_{\boldsymbol{Tab}}\mathrm{d}\boldsymbol{n}_{\boldsymbol{b}}, \tag{31}$$

where the tangent stiffness matrix is:

$$\boldsymbol{K}_{\mathrm{T}} = \left(\boldsymbol{C}^T\widehat{\boldsymbol{S}^T\boldsymbol{x}_c}\boldsymbol{C}\right)\otimes\boldsymbol{I}_3 + \boldsymbol{A}_{1c}\widehat{\boldsymbol{E}}_{tc}\widehat{\boldsymbol{A}}_c\hat{\boldsymbol{l}}_c^{\ -3}\boldsymbol{A}_{1c}^{\mathrm{T}}, \tag{32}$$

$$\boldsymbol{A}_{1c} = (\boldsymbol{C}^T\otimes\boldsymbol{I}_3)b.d.(\boldsymbol{H})\boldsymbol{S}^T, \tag{33}$$

$$\boldsymbol{K}_{Taa} = \boldsymbol{E}_a^{\mathrm{T}}\boldsymbol{K}_{\mathrm{T}}\boldsymbol{E}_a, \tag{34}$$

where $b.d.(\cdot)$ is an operator that converts a vector into a block diagonal matrix. By setting the damping matrix $\boldsymbol{D} = \boldsymbol{0}$, external force $\boldsymbol{f}_{ex} = \boldsymbol{0}$, and $\boldsymbol{n}_b = \dot{\boldsymbol{n}}_b = \ddot{\boldsymbol{n}}_b = 0$, we get the free vibration response of the CTS:

$$\boldsymbol{M}_{aa}\mathrm{d}\ddot{\boldsymbol{n}}_a + \boldsymbol{K}_{Taa}\mathrm{d}\boldsymbol{n}_a = 0, \tag{35}$$

The solution to the homogeneous equation has the form: $\mathrm{d}\boldsymbol{n}_a = \boldsymbol{\phi}\,sin(\omega t - \theta)$, where $\omega$ and $\boldsymbol{\phi}$ are natural frequency and corresponding modes. Substitute the solution to Eq. (35), we get:

$$(\boldsymbol{K}_{Taa} - \omega^2\boldsymbol{M}_{aa})\boldsymbol{\phi}\sin(\omega t - \theta) = \boldsymbol{0}. \tag{36}$$

Since $sin(\omega t - \theta)$ is not always zero, we have the standard eigenvalue problem:

$$\boldsymbol{K}_{Taa}\boldsymbol{\phi} = \omega^2\boldsymbol{M}_{aa}\boldsymbol{\phi}. \tag{37}$$

### 2.3. Statics equation of CTS with constraints

The CTS statics can be obtained by neglecting the dynamic term related to acceleration and speed in Eq. (25), written in terms of nodal vector, we have:

$$\boldsymbol{E}_a^T \boldsymbol{K}\boldsymbol{n} = \boldsymbol{w}_a, \tag{38}$$

in which $\boldsymbol{K} = (\boldsymbol{C}^T(\widehat{\boldsymbol{S}^T\boldsymbol{x}_c})\boldsymbol{C}) \otimes \boldsymbol{I}_3$ is the stiffness matrix and $\boldsymbol{x} = \boldsymbol{S}^T\boldsymbol{x}_c$ is the force density vector. The equation can be written in terms of force densities of the structure elements:

$$\boldsymbol{A}_{1ca}\boldsymbol{x}_c = \boldsymbol{w}_a, \tag{39}$$

where $\boldsymbol{A}_{1ca}$ is:

$$\boldsymbol{A}_{1ca} = \boldsymbol{E}_a^T(\boldsymbol{C}^T \otimes \mathbf{I}_3)b.d.(\boldsymbol{H})\boldsymbol{S}^T. \tag{40}$$

Substitute $\boldsymbol{x}_c = \hat{\boldsymbol{l}}_c^{-1}\boldsymbol{t}_c$ into Eq. (40), we can get the equivalent equilibrium equation in terms of the force:

$$\boldsymbol{A}_{2ca}\boldsymbol{t}_c = \boldsymbol{w}_a, \tag{41}$$

where $\boldsymbol{A}_{2ca}$ is:

$$\boldsymbol{A}_{c2a} = \boldsymbol{E}_a^T(\boldsymbol{C}^T \otimes \mathbf{I}_3)b.d.(\boldsymbol{H})\boldsymbol{S}^T\hat{\boldsymbol{l}}_c^{-1}. \tag{42}$$

## 3. Design of a deployable cable dome based on the CTS

In this section, a deployable cable dome based on the CTS and Levy dome topology is proposed by clustering certain strings in a traditional Levy dome. The topology and configuration of a Levy dome are first introduced. Then, to make the Levy dome deployable with structure global stability guaranteed, the arrangement of clustered strings in a CTS Levy dome is designed and studied. Followed by the description of the deployment trajectory.

### 3.1 Topology and configuration

Compared with other cable dome forms such as Geiger, Kiewitt, Hybrid [33], the Levy dome has a better advantage to be modified into a CTS because all strings in the same group are connected [4]. By changing the rest length of clustered strings, the configuration and prestress can be easily adjusted. The number of actuators is greatly reduced because only one actuator is needed in each clustered string.

A typical Levy dome, shown in Figure 1, consists of nine groups of members: outer bar (OB), inner bar (IB), outer ridge strings (ORS), outer diagonal strings (ODS), inner ridge strings (IRS), inner diagonal strings (IDS), outer hoop strings (OHS), inner hoop strings (IHS), and top hoop strings (THS). Levy dome consists of five groups of nodes: outer top nodes (OTN), outer bottom nodes (OBN), inner top nodes (ITN), inner bottom nodes (IBN), and pinned nodes (PN).

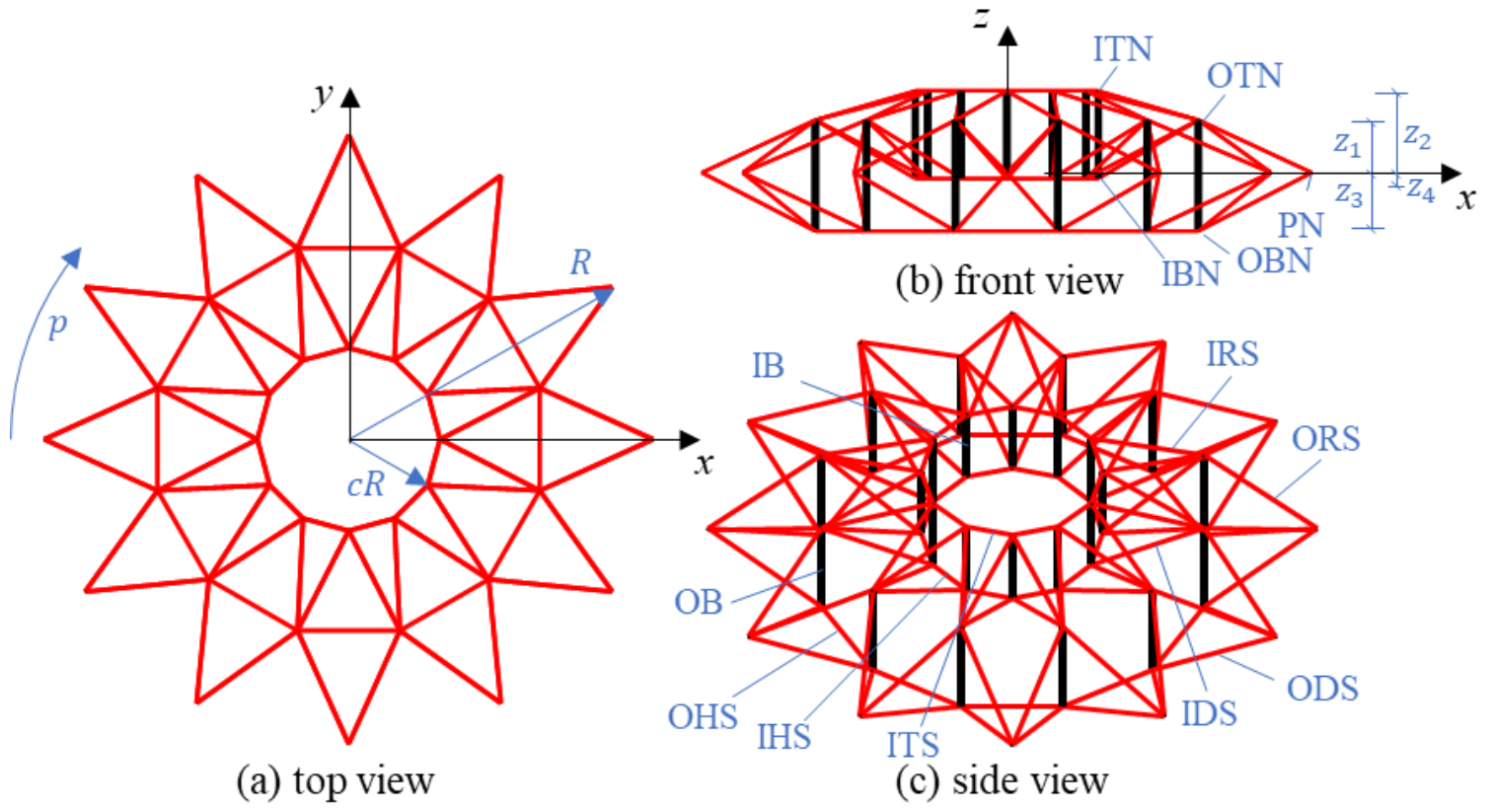

Figure 1: Configuration of a Levy cable dome. Black and red and lines are bars and strings.

The cable dome is rotationally symmetric about the Z-axis, and the rotation angle is $2\pi/p$, where $p$ is the structure complexity of the cable dome in a circular direction. The radius of the circumcircle of the pinned nodes is $R$, the radius of the inner hoop and outer hoop are $cR$ and $(c+1)R/2$, where $c$ is the ratio of the inner ring to the outer ring. The Z-coordinate of the OTN, OBN, ITN, IBN are $z_1$, $z_2$, $z_3$, and $z_4$. Based on the defined parameters, the configuration of a Levy cable dome can be parameterized by several variables: radius of the outer ring ($R$), deployment ratio ($c$), complexity ($p$), Z-coordinate of the free nodes $z_1$, $z_2$, $z_3$, and $z_4$. So, for a given complexity $p$ of the cable dome, for bars and strings, there will be $p$ OS, $p$ IS, $2p$ ORC, $2p$ ODC, $2p$ IRC, $2p$ IDC, $p$ OHC, $p$ IHC, $p$ THC; for nodes, there will be $p$ OTN, $p$ OBN, $p$ ITN, $p$ IBN.

Due to the rotationally symmetric property, the nodes of a Levy dome can be divided into $p$ units. Using Cartesian coordinates in an inertially fixed frame, the nodal coordinate matrix of PN, OTN, OBN, ITN, and IBN in the first unit is:

$$\boldsymbol{N}_{u,1} = \begin{bmatrix} R & r_o\cos\beta & r_o\cos\beta & cR & cR \\ 0 & r_o\sin\beta & r_o\sin\beta & 0 & 0 \\ 0 & z_1 & z_2 & z_3 & z_4 \end{bmatrix}, \tag{43}$$

where $r_o = \frac{(c+1)R}{2}$ is the radius of the outer hoop strings, and $\beta = \pi/p$ is half of the rotational angle between the units. The nodal coordinate matrix of the $i$th units $\boldsymbol{N}_{u,i}$ can be calculated by rotating the first unit $\boldsymbol{N}_{u,1}$ along the Z-axis:

$$\boldsymbol{N}_{u,i} = \boldsymbol{T}^{i-1}\boldsymbol{N}_{u,1}, \tag{44}$$

where the rotation matrix $\boldsymbol{T}$ is:

$$\boldsymbol{T} = \begin{bmatrix} \cos 2\beta & -\sin 2\beta & 0 \\ \sin 2\beta & \cos 2\beta & 0 \\ 0 & 0 & 1 \end{bmatrix}. \tag{45}$$

With the topology shown in Figure 1, one can write the $C_b$ and $C_s$ for bars and strings, a simple approach can be found in [34].

### 3.2 Arrangement of clustered strings

For a deployable dome, one can build it with individual strings, but this will yield a large number of actuators. And the hardware implementation, communication, and control of all the actuators would be a complicated problem. In addition, the cost of the structure would be high. However, it is possible to reduce the number of actuators by using one actuator to drive a clustered string (composed of a set of connected adjacent strings). Indeed, the most direct way to modify a traditional Levy cable dome into a deployable cable dome is to replace all strings in the same group into a clustered string. But there can be global stability problems with the structure if clustered strings are arranged in this way. Thus, the proper way to cluster strings to guarantee the stability of the structure and reduce the actuator number is our concern in this subsection.

We choose a classical Levy dome with complexity $p$=12 as an example. The strings in one group are equally divided into $n_c$ clustered strings. Both ends of the clustered string should be pinned on joints, while the intermediate part of the clustered string can pass over the pulleys. This means $p$ is divisible by $n_c$. For $p$=12, $n_c$ can be 1, 2, 3, 4, 6, and 12. The radius of the outer ring is $R = 50\text{m}$. For the simplification of comparison, the deployment ratio is set to $c = 0.3$, and the Z-coordinate of vertical bars are $z_1 = 8.663\text{m}$, $z_2 = 13.458\text{m}$, $z_3 = -9.623\text{m}$, $z_4 = -0.960\text{m}$ (the coordinates are determined by a standard arc, whose arch rise is 15% length of its span). The prestress of the inner bar (IB) is assigned to be $-5{,}000\text{N}$, so the prestress of the structure can be determined by the equilibrium. The cross-sectional area of strings and bars is designed to be 10% of their yielding or buckling stress. The mass of the bars and strings are calculated by the minimal mass algorithm developed in [35].

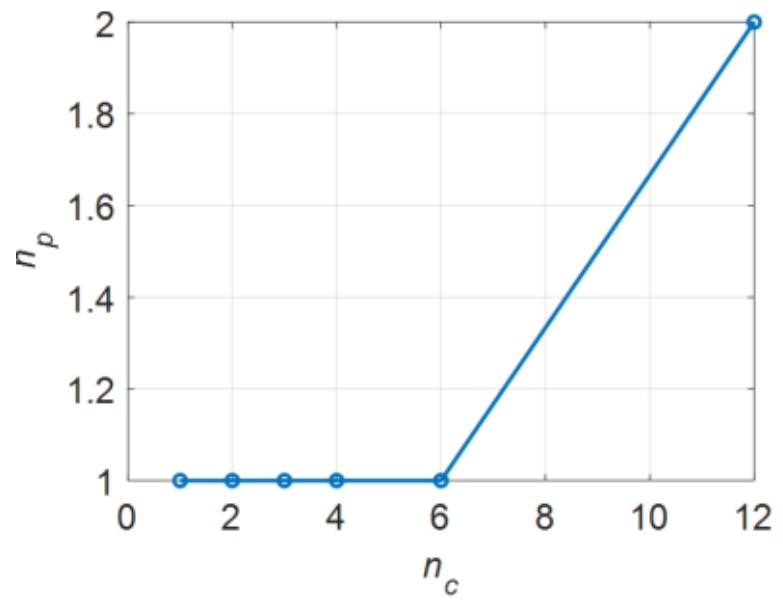

Figure 2. Curve of $n_p$ vs. $n_c$.

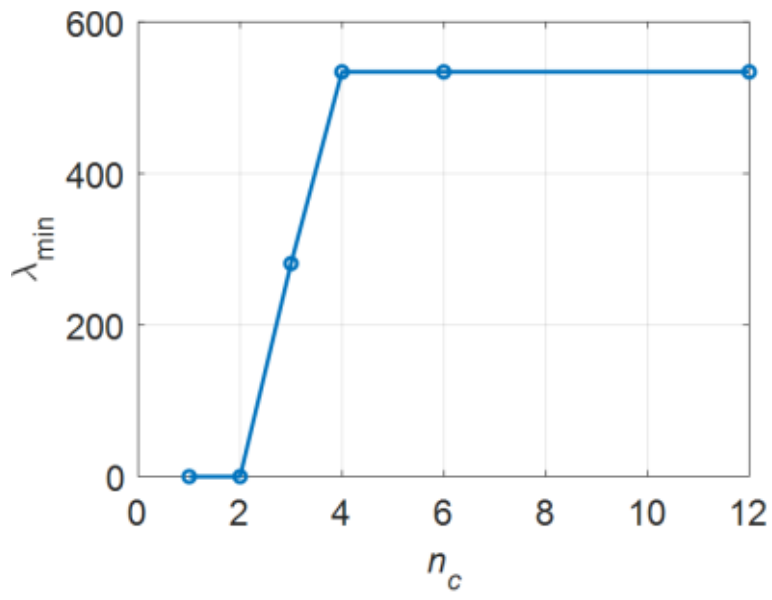

Figure 3. Curve of d $\lambda_{min}$ vs. $n_c$.

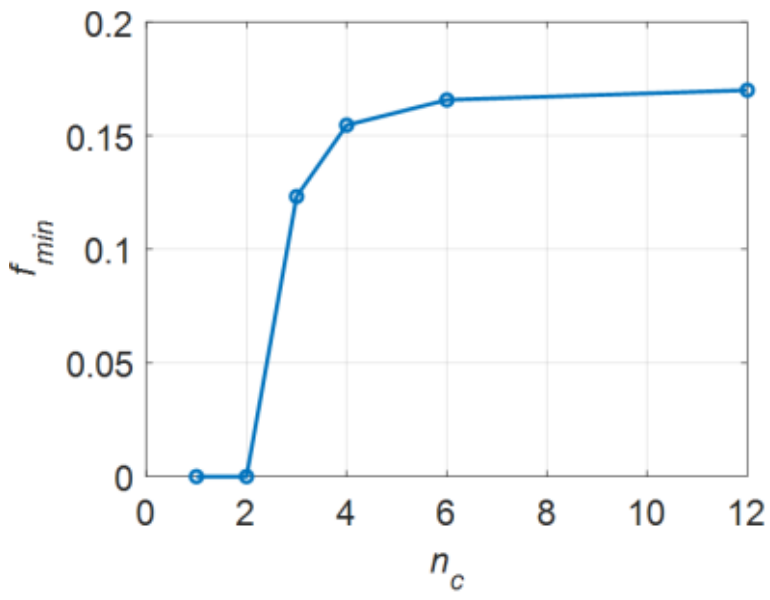

Figure 4. Curve of $f_{min}$ vs. $n_c$.

For each feasible $n_c$, singular value decomposition of the equilibrium matrix in Eq. (40) is conducted, and the number of prestress mode $n_p$ is examined, shown in Figure 2. We can see that the number of prestress mode is 2 if $n_c = 12$, which means all the strings are individual ones. Results also show that the number of prestress mode is 1 if a clustered string is used in every group of strings.

The eigenvalue analysis of the tangent stiffness matrix is conducted to check the structure's global stability, and the minimal eigenvalue $\lambda_{min}$ of the tangent stiffness matrix is shown in Figure 3. If 1 or 2 clustered strings are used in each string group, there will be a global stability problem. To guarantee the structure is globally stable, there should be more than 2 clustered strings used in each group of strings. The minimal eigenvalue of the tangent stiffness matrix is the same for $n_c = 4, 6, 12$. We also analyzed the free vibration, and the natural frequency is obtained by the generalized eigenvalues of the tangent stiffness matrix and mass matrix according to Eq. (37). The minimal natural frequency $f_{min}$ in each $n_c$ is shown in Figure 4, and we can see that the structure has no stiffness if $n_c = 1, 2$ and the minimal natural frequency increases with $n_c$.

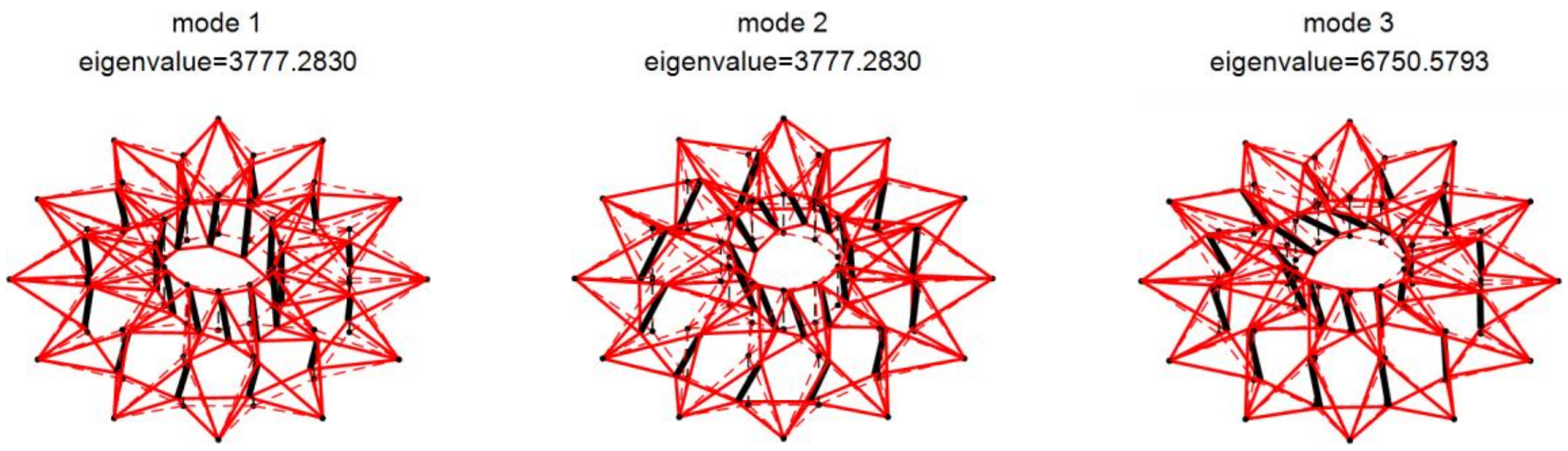

Figure 5 Eigenvalue (N/m) and eigenvectors of the tangent stiffness matrix of the cable dome.

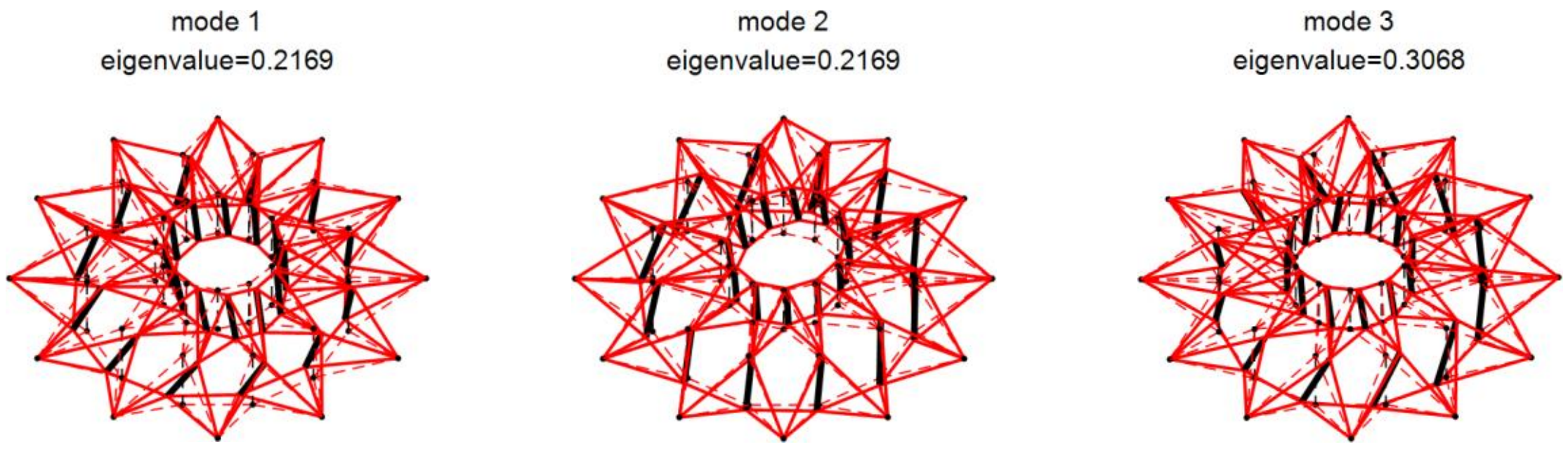

Figure 6 Natural frequency (Hz) and first three free vibration modes of the cable dome.

Under the global stability constraints, the number of clustered strings in each group should be minimized to reduce the expense of the structure. So we use three clustered string, $n_c = 3$, in each group of the Levy dome. The mode shapes corresponding to the three minimal eigenvalues of the tangent stiffness matrix is shown in Figure 5, and the first three free vibration mode is shown in Figure 6.

### 3.3 Deployment trajectories

Since the CTS cable dome is symmetric and several variables can parameterize its configuration, the structure can be easily deployed by changing one variable while keeping the others constant. In this section, the deployment trajectory of the CTS cable dome is obtained by changing the coefficient $c$. As $c$ increases from 0 to 1, the structure moves from folded configuration to deployed configuration, shown in Figure 7. Since the structure is not stable for $c > 0.95$. For the safety consideration of the CTS cable dome in engineering practice, we deploy it only in the interval of $c \in [0,0.9]$.

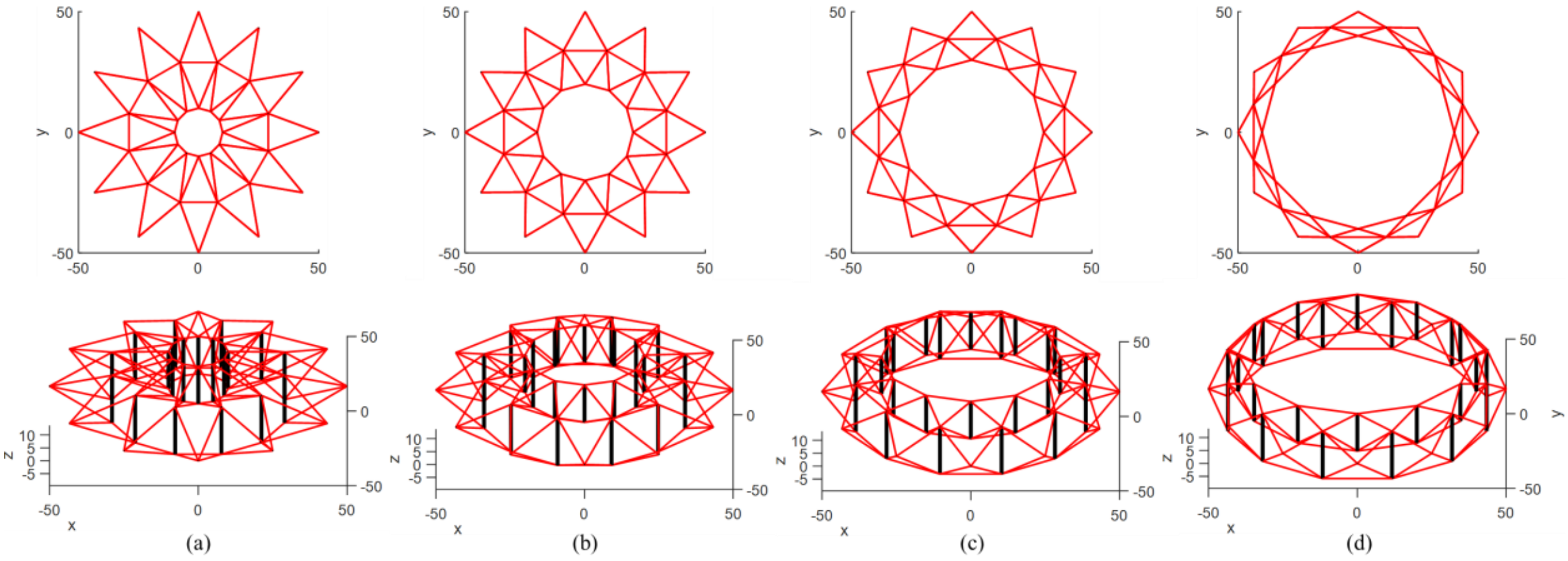

Figure 7 Deployment trajectories of the CTS cable dome, (a) $c = 0.2$, (b) $c = 0.4$, (c) $c = 0.6$, (d) $c = 0.8$.

The CTS cable dome always has one prestress mode on the trajectory. The mechanical properties of the cable dome on the deployment trajectory, including the number of self-stress modes $n_p$, the minimal eigenvalue of the tangent stiffness matrix $\lambda_{min}$ and minimal natural frequency $f_{min}$, should be examined to ensure the feasibility of the trajectory, and the results are shown in Figure 8 - Figure 10.

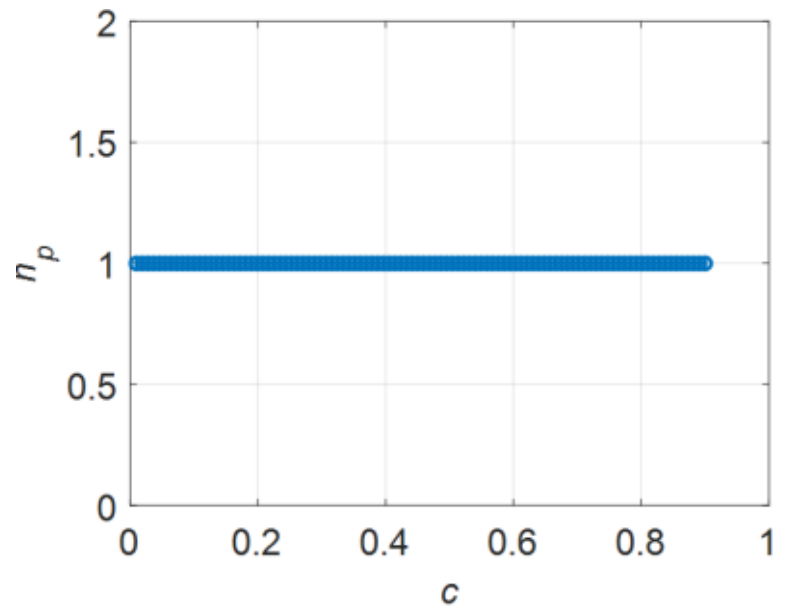

Figure 8. Curve of $n_p$ vs. deployment ratio $c$.

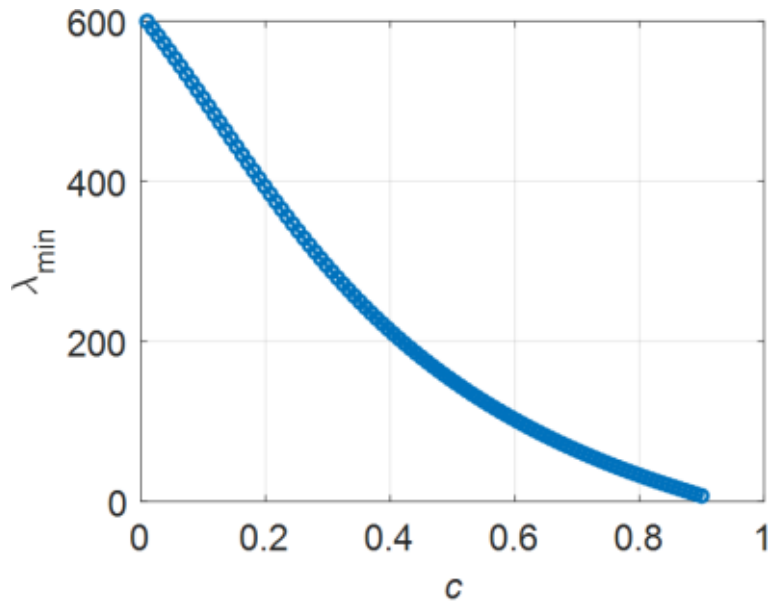

Figure 9. Curve of $\lambda_{min}$ vs. deployment ratio $c$.

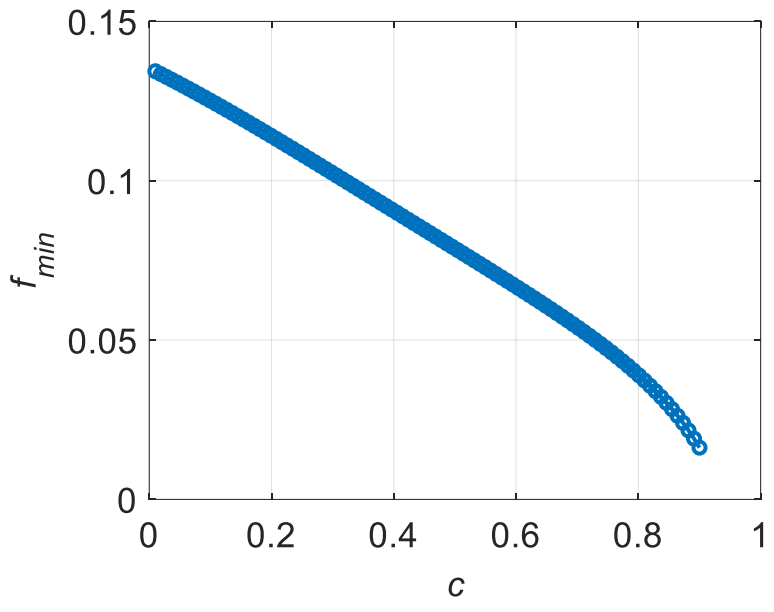

Figure 10. Curve of $f_{min}$ vs. deployment ratio $c$.

## 4. Deployment analysis of the deployable cable dome

In this section, the actuation strategy of the CTS cable dome is proposed based on the quasi-static analysis of the deployment process. Then, the inertia force of members is considered, and the dynamic analysis of the deployment is conducted at different actuation speeds.

### 4.1 Quasi-static deployable analysis

In section 3.3, we propose a deployment trajectory by simply changing one shape parameter $c$. The prestress of all the members in the cable dome is shown in Figure 11. And the rest length of clustered strings in different groups is shown in Figure 12.

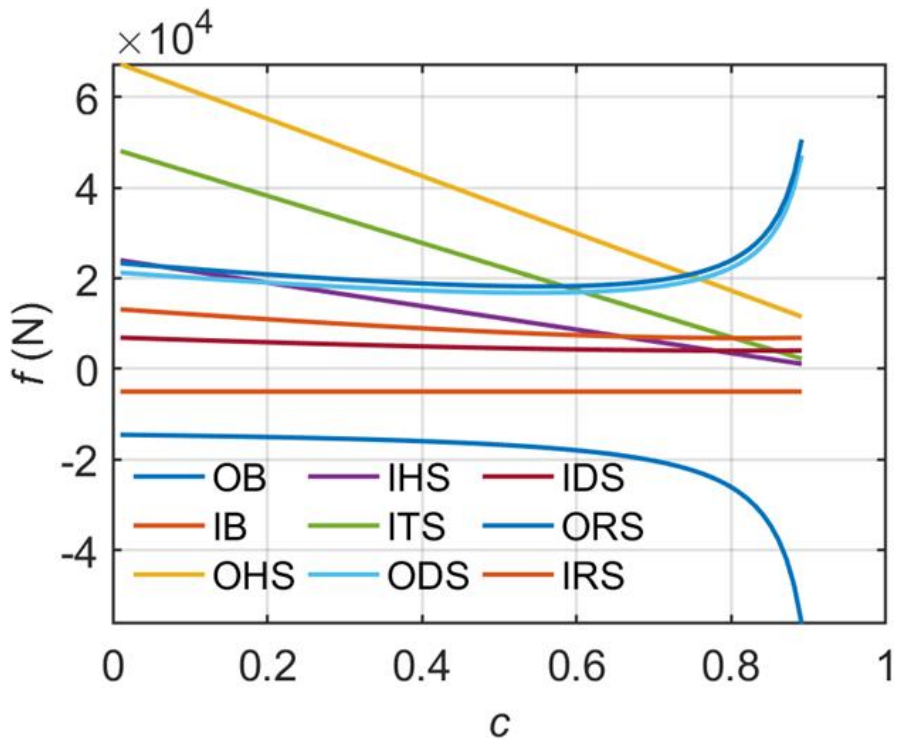

Figure 11 Member prestress in groups vs. deployment ratio $c$.

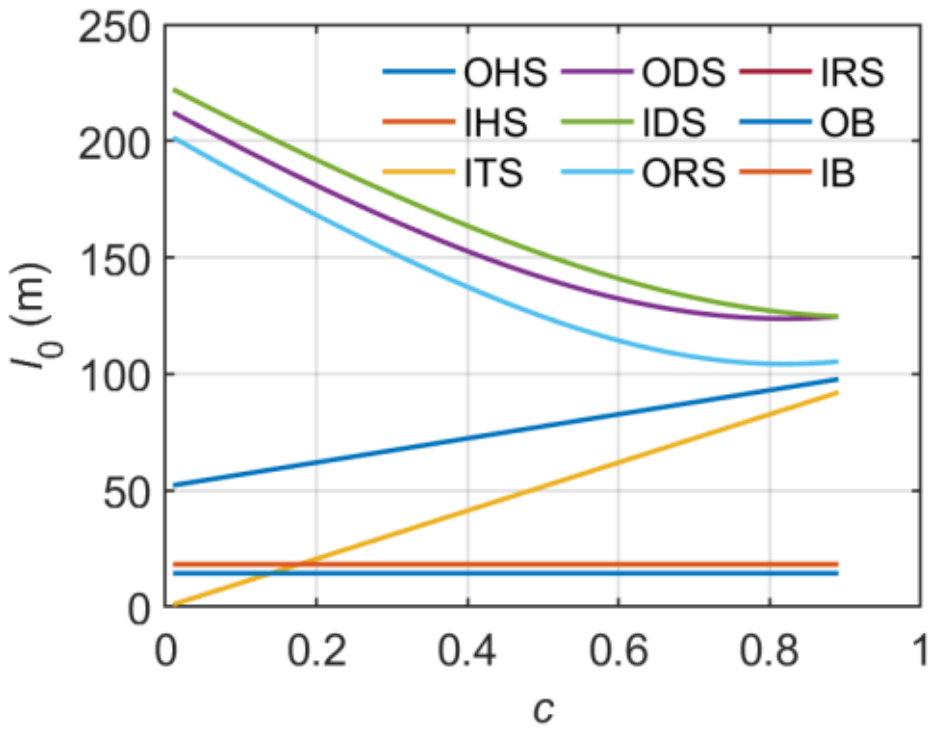

Figure 12 Rest length of the clustered strings vs. deployment ratio $c$.

As we can see, for the cable dome deploys from and initial configuration to the final configuration, the rest length of bars is constant, the rest length of OHS, HIS, ITS is increasing, while that of ODS, IDS, ORS, and IRS is decreasing.

## 4.2 Dynamic deployment analysis

In this subsection, we study the dynamic deployment performance of the CTS cable dome. We use steel bars and aluminum strings, the Young's modulus are $2.06 \times 10^{11}$Pa and $6 \times 10^{10}$Pa. The material densities of the bars and strings are 7,870 kg/m$^3$ and 2,700 kg/m$^3$. In the numerical simulation, the density is increased to 50 times its real value to compensate for the effect of the additional mass of joints and actuators. This modification is introduced to exaggerate the inertia force of members because the members' mass by the minimal mass design is relatively small.

At the initial state, the structure is static, the deployment ratio $c = 0.2$ and the prestress and rest length of members are applied based on the equilibrium analysis. The gravity force is negligible compared with the axial force of members, so the self-weight is not considered. Then, the cable dome is actuated by changing the rest length of the strings. In the deployment process, the actuation strategy of each clustered string is given in Figure 12. The rest length of OHS and IHS is increasing while the rest length of the other strings is decreasing progressively. For the folding process, the reverse procedure is also employed.

The dynamic analysis of deployment is investigated at different actuation speeds. The analysis time is set to be 1, 2, and 4s. The rest length of clustered strings is changed according to the deployment ratio $c$ from 0.2 to 0.8 in the first half period of the total time, and the time interval beyond the actuation time is used to observe the vibration characteristics of the dynamical system. For all the simulations, the time step sizes are fixed to be 0.001s. The damping ratio is d = 0.01 for the metal material.

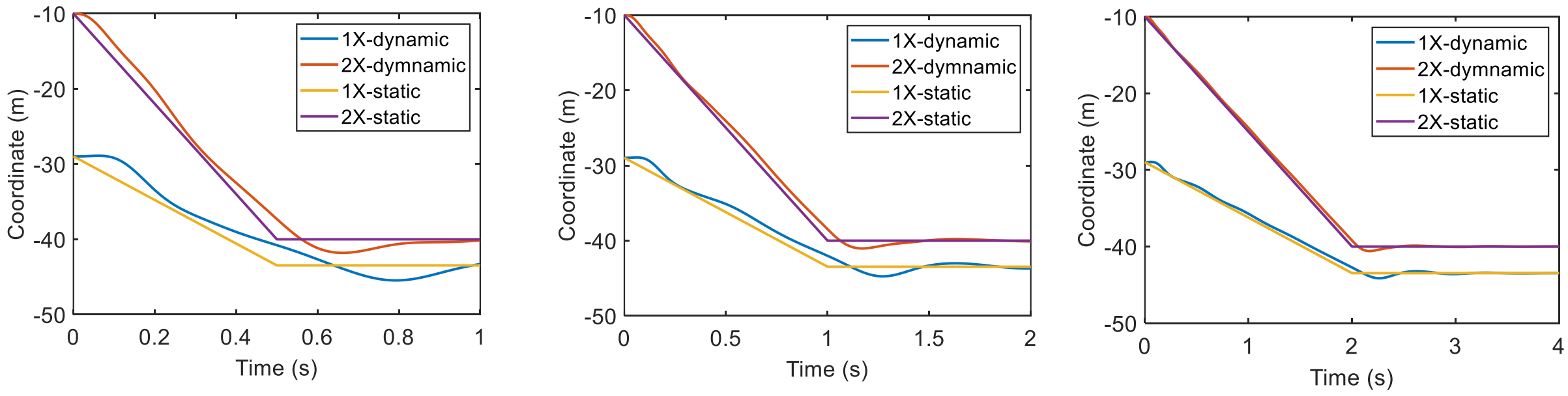

(a) (b) (c)

Figure 13. Time history curves of the nodal coordinates, (a) t=1s, (b) t=2s, (c) t=4s.

Figure 13 gives the time history curves of the X nodal coordinate of the OTN (1X) and ITN (2X). The corresponding static solutions are also presented as comparisons. The dynamic curve differs from each other. For example, at the relatively low actuation speeds (t = 4s), the dynamic solutions are very close to the static solution. The structure can be actuated to the target configuration with a small vibration. The amplitude of vibration increases if the actuation speed increases. For high actuation speed (t = 1s, 2s), the deviation of dynamics curves from static curves are much noticeable since higher actuation speed results in higher acceleration and inertia force.

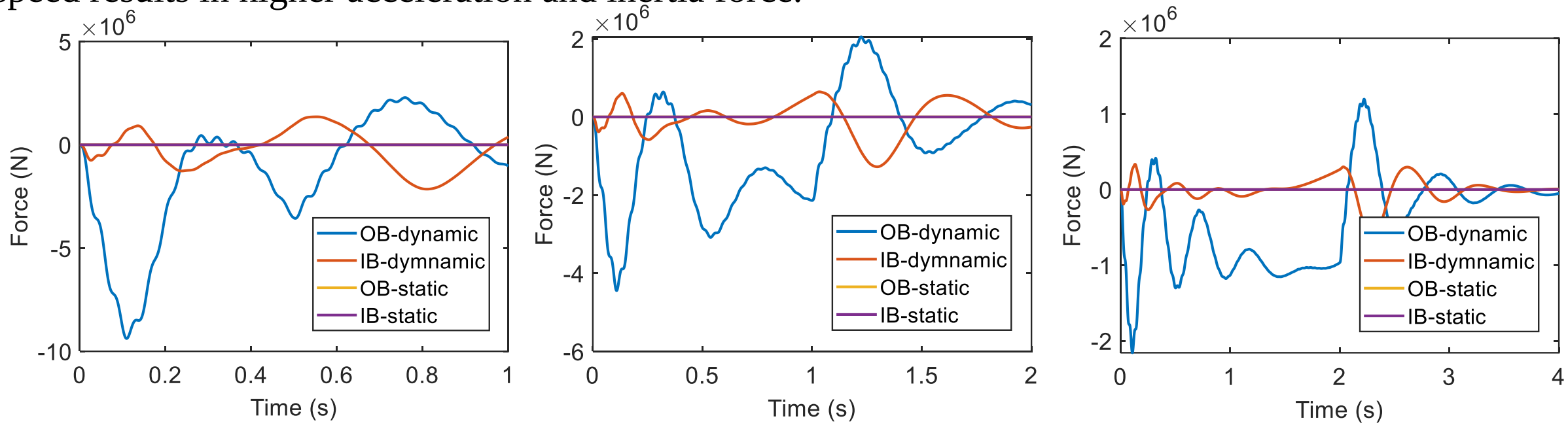

(a) (b) (c)

Figure 14. Time history curves of the member force, (a) t=1s, (b) t=2s, (c) t=4s.

Figure 14 gives the time history curves of the member forces of OB and IB, and we can observe that the member force decreases with a longer actuation time. According to the static analysis, we know the force of bars is around 5,000N. However, the member force of bars in the dynamic analysis is around 100 times higher than the static results. That is to say, the effect of the dynamics, and the axial stiffness of the clustered strings is very high. Even a slight change of rest length in each time step will lead to high stress in the structure members.

## 5. Conclusions

This paper proposed the design of a deployable cable dome based on the combination of the traditional Levy dome and clustered strings. To analysis the proposed design, we presented the explicit statics (three forms in terms of nodal coordinate, force density, and force vectors) and dynamics equations with any boundary conditions of CTS. The statics analysis guided us in choosing the clustered strings such that the existence of prestress mode and global stability of the structure is guaranteed. Results show that at least two clustered strings in each group of strings of the deployable cable dome are required to meet the stability constraint. Then, we studied the deployment of clustered tensegrity both quasi-statically and dynamically by varying one shape parameters while keeping the other parameters constant. The quasi-static analysis of the deployment process is studied and compared with the dynamic one with different actuation speeds. The results show that the dynamic effects must be considered for fast actuation speeds to avoid large overshot and significant stress changes in the structure members. The approaches developed in this paper can also be used to design and analyze various kinds of CTS.